# MONOLAYER CHARACTERISTICS OF PYRENE MIXED WITH STEARIC ACID AT THE AIR-WATER INTERFACE

MD. N. ISLAM$^\$$, D. BHATTACHARJEE$^\$$, SYED ARSHAD HUSSAIN$^{\$\S}$*

$\$$ Department of Physics, Tripura University, Suryamaninagar-799130, Tripura India
§ Centrum voor Oppervlaktechemie en Katalyse, Katholieke Universiteit Leuven, Kasteelpark, Arenberg 23, 3001 Leuven, Belgium.



**Abstract:**
In the present communication we report the monolayer characteristics of pyrene mixed with stearic acid (SA) at the air-water interface. The monolayer properties are investigated by recording and analyzing the surface pressure-area per molecule isotherm ($\pi - A$) of the pyrene-SA mixed films. It is observed that the pyrene and SA are miscible in the mixed monolayer. This miscibility/nonideality leads to phase separation between the constituent components (pyrene and SA). BAM image of the mixed monolayer confirms the miscibility or nonideal mixing at the mixed monolayer.

* Email: sa_h153@hotmail.com


## 1. Introduction

In recent years monolayers which often regarded as two dimensional model systems, have been the subject of extensive studies. The main interest has been focused on phase transitions and ordering process into two dimensions[1,2]. The Langmuir-Blodgett (LB) technique seems very promising in comparison with the other contemporary monolayer forming techniques. Because here the surface pressure-area per molecule ($\pi - A$) isotherm reflects the intermolecular forces operating in two dimensional arrangements of molecules and these data provide information on the molecular packing in two dimensions. By extrapolation of the isotherm from the solid phase (steep rising region) to zero surface pressure, the size of the molecules can be estimated assuming a dense packing of these molecules in two dimensions[3].

Direct visual observations of the textures of Langmuir monolayer become possible few years ago, first by the means of fluorescence imaging microscopy and most recently caused by the new technique of Brewster-angle-microscopy (BAM). Detailed description of BAM technique and principle has been reported in several publications[4,5]. The main advantage of BAM with respect to fluorescence imaging microscopy is that using BAM one can directly visualize the morphology and domain structure of a monolayer at air-water interface without adding any further fluorophore molecules.

Amphiphilic molecules are ideally suitable for LB technique. However investigation suggest that nonamphiphilic molecules can also form well organized stable Langmuir and Langmuir-Blodgett (LB) films when they are mixed with suitable matrix molecules viz. a long chain fatty acid (stearic acid or arachhidic acid) or an inert polymer (polymethyl methacrylate or polystyrene)[6].

Recently mixed monolayers have gained more and more interest in the field of monolayer research due to the vast properties of the multicomponent system then that of the pure one[6-8]. For some cases, to obtain good stability and transfer ratio while transferring to solid substrate it is required to add a second component in the monolayer[6-8]. For mixed monolayer it is important to have idea about the mixing behavior of the constituent molecules to optimize the deposition parameters to have a good quality LB films for specific applications.

In the present paper, we report details monolayer characteristics of pyrene mixed with stearic acid at the air-water interface. Pyrene is well known due to intense fluorescence with long life time



and strong excimer emission and widely used as phohtophysical probe [9-11]. Although pyrene and its derivatives are extensively studied in the restricted geometry of LB film [9-11]. However details investigations of the monolayer characteristics of pyrene at the air-water interface has never been reported before.

## 2. Experimental
### 2.1. *Materials*
Pyrene (99% pure) purchased from Aldrich Chemical Co, USA was purified under vacuum sublimation followed by repeated re-crystallization before use. SA (purity > 99%) purchased from Sigma Chemical Company. Spectroscopic grade Chloroform (SRL, India) was used as solvent and its purity was checked by fluorescence spectroscopy before use. Solutions of pyrene, SA as well as pyrene-SA mixture at different mole fractions were prepared in chloroform solvent and were spread on the water surface of the LB trough.

### 2.2. *Isotherm measurement*
A commercially available Langmuir-Blodgett (LB) film deposition instrument (Apex-2000C, India) was used for isotherm measurement. A surface pressure-area per molecule isotherm was obtained by a continuous compression of a monolayer at the air-water interface of the LB trough by a barrier system. The surface pressure at the air-water interface was measured using wilhelmy plate arrangement attached to a microbalance, whose output was interfaced to a microcomputer, which control the barrier movement confining the monolayer at the air-water interface. Milli-Q water was used as subphase and the temperature was maintained at $24^0$C.

Before each isotherm measurement, the trough and barrier were cleaned with ethanol and then rinsed by Milli-Q water. In addition the glass ware was cleaned prior to the use. The surface pressure fluctuation was estimated to be less than 0.5 mN/m during the compression of the entire trough surface area. Then the barrier was moved back to its initial position and the sample containing monolayer forming material was spread on the subphase using a Hamilton microsyringe. After a delay of 30 minutes, to evaporate the solvent, the film at the air water interface was compressed slowly at the rate of 5 mm/min to obtain a single surface pressure versus area per molecule (π-A) isotherm. All isotherms were run several times with freshly prepared solution.

### 2.3. *Brewster angle microscopy (BAM) imaging*
BAM imaging was recorded using an NFT (Gö̈ttingen, Germany) Mini BAM, aligned perpendicular to the direction of compression of the pyrene-SA (0.5 M of pyrene) mixed monolayer. The polarizer and analyzer were set to p-polarization, and the incoming laser light (688 nm, 30 mW) was limited to an angle of incidence of 52–54° (Brewster angle for aqueous sub phase). Image was captured by a low geometrical distortion, sensitive, black-white frame transfer CCD camera were transferred in real time to a computer for analysis and use. The images were acquired and analyzed with the Image-Pro Plus 2.1 software (Media Cybernetics, Silver Spring, MD) for maximum resolution.

## 3. Results and discussions
### 3.1. *Monolayer characteristics at the air-water interface*
To study the monolayer characteristics of pure pyrene at the air-water interface, 100 $\mu l$ of dilute chloroform solution of pyrene ($0.5 \times 10^{-3}$ M) was spread at the air-water interface by a micro syringe. After allowing 30 minute to evaporate the solvent, the barrier was compressed very slowly at a speed of 10 mm/min to record the surface pressure versus area per molecule $(\pi - A)$ isotherm. It was observed (figure 1) that the surface pressure did not rise beyond 9.45 mN/m and no distinct phases were observed in the isotherm curve. Addition of larger amount of solution resulted in the formation of microcrystalline domains, which were visible even to the naked eye. Moreover, the islets once formed as a result of barrier compression did not disintegrate at the molecular level upon relaxation of surface pressure by expanding the barrier rather remained as smaller islets. Also repeated attempt to transfer this monolayer onto quartz substrate were failed. However, pyrene, when mixed with stearic acid it was observed that stable, compressible, self supporting films were formed at the air-water interface, which can be easily transferred onto quartz substrate to form stable mono- or multilayer LB films.

Figure 1 shows the $\pi - A$ isotherms of pyrene-SA mixed films at the air-water interface for different mole fractions of pyrene along with the pure pyrene and SA isotherm.

The pure SA isotherm is a smoothly rising curve with a lift off area 0.255 nm$^2$ and collapsed at about 53.37 mN/m. A phase transition point occurred at about 22.9 mN/m indicating the attainment of solid phase, with a limiting area (estimated by extrapolating the constant slope region of the isotherm to zero surface pressure) of about 0.209 nm$^2$/monomer. Also the area per molecule of pure SA were 0.225 nm$^2$ and 0.213 nm$^2$ at surface pressures 15 and 25 mN/m respectively. These values and the



shape of the pure SA isotherm were in well agreement with the reported results [8,12].

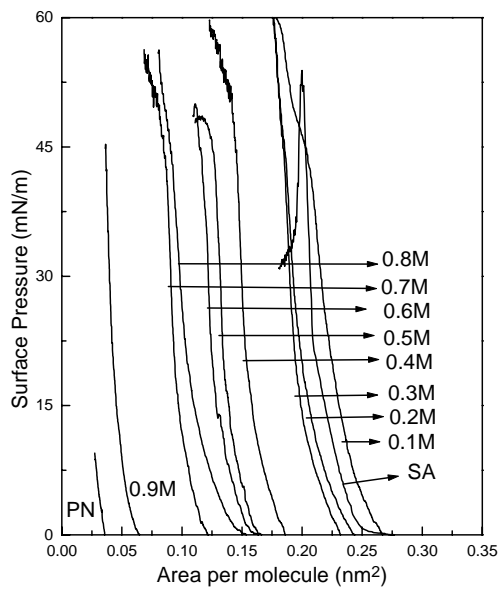

*Fig. 1. Surface pressure (π) vs. area per molecule (A) isotherms of pyrene in SA matrix at different mole fractions of pyrene along with pure pyrene and SA isotherm. The numbers denote corresponding mole fractions of pyrene in SA matrix. SA and PN corresponds to the pure SA and pyrene isotherm respectively.*

It was observed that the area per molecule for pyrene-SA mixed isotherm was lower than the pure SA isotherm except 0.1 M of pyrene in SA. Also the area per molecule gradually decreases with the increase in mole fraction of pyrene. This decrease in area per molecule with the increasing mole fraction of pyerene may be due to either loss of pyrene molecules through precipitation in the bulk surfaces or the accommodation of pyrene molecules in between the fatty acid (SA) chains. To confirm this small amount of water from just below the air-water interface were sucked out by a bent tube [13] and the fluorescence of the water sample was checked. It was confirmed from the failure to detect any fluorescence that the pyrene molecules were not lost through submerging the water sub phase. Therefore the most plausible explanation may be that the pyrene molecules are pushed up in between SA chains in such a way that very small area of the pyrene molecule occupy at the air-water interface. Figure 2 shows the schematic of pyrene-SA monolayer at the air-water interface.

To have more information about the mixed monolayer at the air-water interface, the compressibility (C) of the mixed films were calculated, which is defined as [14]

$$C = -\frac{1}{a_1}\frac{a_2 - a_1}{\pi_2 - \pi_1}$$

Where $a_1$ and $a_2$ are the area per molecules at the surface pressures $\pi_1$ and $\pi_2$ respectively. In the present case the values of $\pi_1$ and $\pi_2$ are chosen as 10 and 30 mN/m respectively.

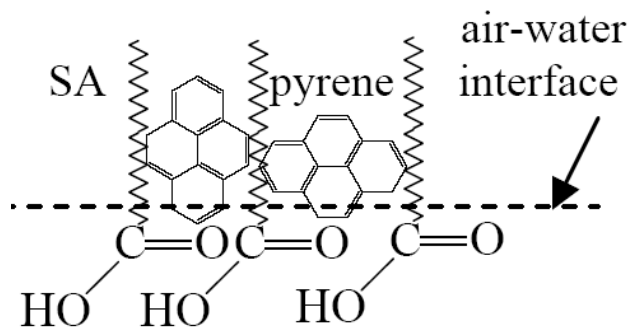

*Fig. 2. Schematic of pyrene-SA monolayer at the air-water interface.*

The basic $\pi - A$ isotherm features of the pyrene-SA mixed films are tabulated in table 1. Although the pure pyrene isotherm did not rise up to 30 mN/m, however, the value of area per molecule of pure pyrene monolayer at 30 mN/m were obtained by extrapolating the pure pyrene isotherm. The compressibility of pure pyrene was found to be 17.39 mN$^{-1}$. In this context it is interesting to note that the isotherms of the films having high compressibility values possess different phases and some times platue like regions and these films posses less stability [15]. So in the present case the higher value of compressibility of pyrene film suggest that pure pyrene film at the air-water interface is unstable. On the other hand compressibility of pure SA isotherm was found to be 5.69 mN$^{-1}$. This is reasonable as fatty acid monolayer on water surface has the closest molecular packing and can be considered as a two dimensional crystalline solid [6]. The compressibility values of pyerene-SA mixed films were also found to be lower than pure pyrene film and closer to the SA film. This was an indicative that pyrene-SA mixed films at the air-water interface were compact enough to form self supporting stable monolayer films.

The collapse pressures and phase transition as well as shape of the pyrene-SA mixed isotherms resembles to that of pure SA monolayer, especially for the SA rich monolayer. Such result indicate that pyrene and SA are not ideally mixed via two-dimensional mixing, according to Crisp's law [16], which was also be verified later by the analysis of excess surface area according to phase rule [14].

*Table 1. Monolayer properties taken from isotherm.*

| Mole fraction of pyrene in SA | Lift off area (nm$^2$) | Limiting area ($A_0$) (nm$^2$) | Collapse pressure (mN/m) | $C$ (mN$^{-1}$) |
|---|---|---|---|---|
| 0 | 0.255 | 0.209 | 53.37 | 5.69 |
| 0.1 | 0.26 | 0.239 | 44.51 | 4.64 |
| 0.2 | 0.24 | 0.205 | 45.39 | 4.91 |
| 0.3 | 0.23 | 0.201 | 44.5 | 4.05 |
| 0.4 | 0.17 | 0.16 | 44.2 | 4.43 |
| 0.5 | 0.163 | 0.14 | 42.89 | 4.14 |
| 0.6 | 0.15 | 0.13 | 35.54 | 5.07 |
| 0.7 | 0.12 | 0.11 | 42.42 | 5.65 |
| 0.8 | 0.14 | 0.12 | 42.7 | 8.32 |
| 0.9 | 0.06 | 0.045 | 45.16 | 9.80 |
| 1 | 0.036 | 0.035 | 9.45 | 17.39 |

### 3.2. Miscibility of pyrene-SA mixed monolayer

An ideal mixed monolayer and a completely immiscible monolayer are absolutely different and almost opposite. Actually there exist continuous intermolecular forces between the constituent molecules in a mixed monolayer. The magnitude of these forces determines the mixing behaviors of the mixed monolayer films. In the case of ideally mixed monolayer of components 1 and 2, the intermolecular forces $F_{11} = F_{12} = F_{22}$, whereas, in a completely immiscible monolayer, $F_{11} > F_{12} < F_{22}$, where $F_{ij}$ are the attractive intermolecular forces between molecules of the two constituent components $i$ and $j$ ($i, j = 1, 2$ respectively).

The miscibility of a mixed monolayer can be examined by quantitative analysis of the excess are ($A^E$) of the mixed monolayer at the air-water interface. The excess area can be obtained by comparing the experimentally observed average area per molecule ($A_{12}$) of the mixed monolayer consisting of the components 1 and 2, with that of an ideally mixed monolayer ($A_{id}$) and is given by $A^E = A_{12} - A_{id}$.

In terms of molecular area of the mixed monolayer, the ideal value of the molecular area, $A_{id}$, is calculated from the molar ratios of the two components, $A_{id} = A_1 N_1 + A_2 N_2$, where $A_1$ and $A_2$ are the monomer areas occupied by the pure components and $N_1$ & $N_2$ are the mole fractions of the pure components in the mixture.

Once two components form an ideally mixed monolayer or they are immiscible, the $A_{id}$ will be equal to $A_{12}$ resulting $A^E = 0$. When $A_{12} \neq A_{id}$ i.e. there is an deviation of experimentally observed molecular areas in the mixed film from the values supposed for ideal mixing, then $A^E \neq 0$. And the value of $A^E$ may be considered as the measure of interaction between the mixed components. This is because the $A^E$ value depends on the intermolecular forces between the constituent molecules in the mixed monolayer.

For an ideally mixed monolayer, the plot of $A_{12}$ versus $N_1$ will be a straight line. Any deviation from the straight line ($A^E = A_{12} - A_{id} \neq 0$) indicates miscibility and non-ideality [14,17]. For a mixed monolayer, if attractive intermolecular force exists then $A^E$ will be negative resulting a negative deviation from the ideal characteristics. On the other hand positive deviation ($A^E > 0$) from the ideal behavior indicates strong repulsive interaction between the constituent components of the mixed monolayer.

Figure 3 represents the plot of area per molecule versus mole fraction of pyrene in the pyrene-SA mixed monolayer at different fixed surface pressures of 5, 10, 15, 20, 25, 30, 35, 40 and 45 mN/m.

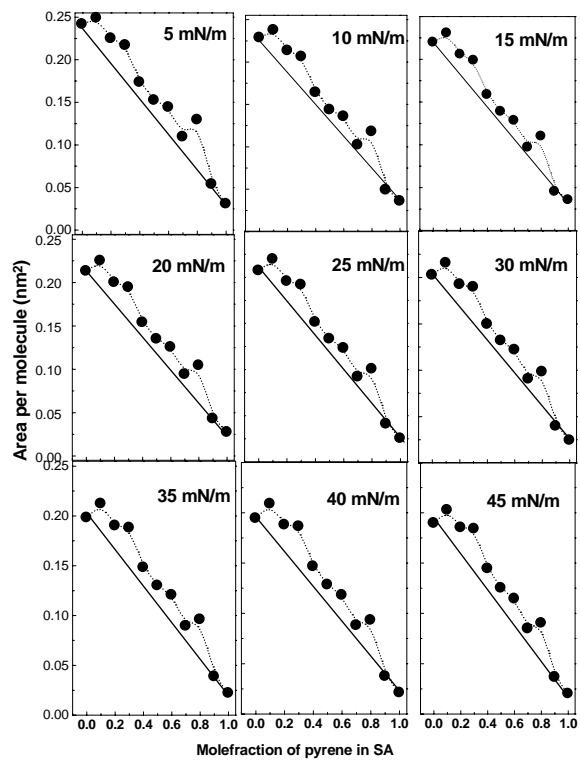

*Fig. 3. The plot of the data of the actual area ($A_{12}$) per molecule versus mole fraction of pyrene in the pyrene-SA mixed monolayer for different fixed surface pressure. The numbers denote the surface pressures.*

From the figures it was observed that the experimentally observed values of area per molecule show positive deviation from the ideal behavior. This indicates that the intermolecular interaction between pyrene and SA is repulsive in nature. Therefore, cohesive forces between the like molecules (pyrene-pyrene & SA-SA) may dominate in the mixed films with respect to the cohesive force between the unlike (pyrene-SA) component. That means that the pyrene and SA are not completely immiscible in the mixed monolayer and do not form ideally mixed monolayer at the air-water interface. This may be due to the fact that pyrene and SA possess different physical and chemical behavior which results some extent of miscibility and non-ideality. The BAM image of the mixed monolayer also support this observation.

Figure 4 represents the BAM image of pyrene-SA (0.5 M) mixed monolayer taken at surface pressures of 10 mN/m. The BAM image reveals distinct circular domains in the mixed monolayer with heterogeneous distribution through out the film. As explained before, this is expected because the chemical and physical nature of pyrene and SA are totally different and they are miscible in the mixed monolayer. Therefore BAM images of pyrene-SA mixed monolayer gives the compelling visual evidence of miscibility / nonideality in the mixed monolayer.

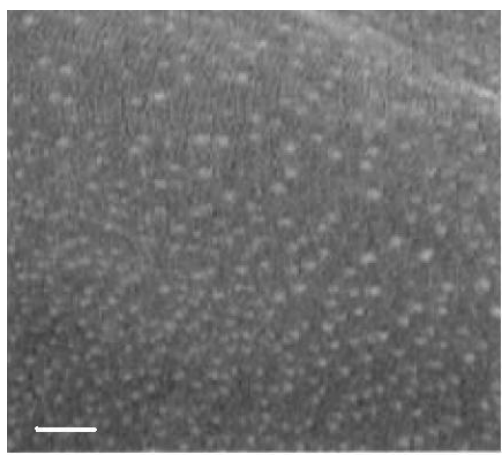

*Fig. 4. BAM image of pyre-SA (0.5 M of pyrene) mixed monolayer taken at surface pressure 10 mN/m. The scale bar represents a length of 250 $\mu m$*

### 4. Conclusion:

We have investigated the monolayer characteristics of pyrene mixed with SA at the air-water interface by Langmuir-Blodgett technique. It has been observed that the pure pyrene does not form stable Langmuir monolayer, however, when they are mixed with SA form self supporting stable monolayer at the air-water interface. The area per molecule of the mixed isotherm systematically decreases with the increase in mole fraction of pyrene in the mixed films indicating successful incorporation of pyrene molecules with in the fatty acid chain. The area per molecule versus mole fraction plot suggest that there exist repulsive interaction between the pyrene and SA and the cohesive force between the like components (SA-SA and pyrene-pyrene) in the mixed monolayer predominates. It was also observed that the pyrene and SA are miscible in the mixed monolayer leading non-ideal mixing. This miscibility leads to the phase separation in the mixed monolayer. BAM image give the compelling visual evidence of miscibility in the mixed monolayer.

**Acknowledgement:**
The authors are grateful to UGC and CSIR, Govt. of India for providing financial assistance through UGC minor project Ref. No. F.1-1/2000(FD-III)/399 and Sanction Ref. No. F.31-30/2005 337 (SR) and CSIR project Ref. No. 03(1080)/06/EMR-II. We are also grateful to H. Leman for BAM image measurement.